\documentclass[aps,prl,notitlepage,superscriptaddress,nofootinbib,tightenlines,twocolumn]{revtex4-1} %
\usepackage{hyperref}
\usepackage{graphicx}
\usepackage{amssymb}
\usepackage{amsmath}
\usepackage{slashed}
\usepackage{bbm} %double stroke letters
\usepackage{turnstile}

\newcommand{\J}{{\cal J}}

\begin{document}

\title{Monte Carlo simulations on the Lefschetz thimble: taming the sign problem}

\author{Marco Cristoforetti}
\affiliation{ECT$^\star$ / FBK, strada delle tabarelle 286 - 38123 - Trento, Italy.}
\affiliation{LISC  / FBK, via sommarive 18 - 38123 - Trento, Italy.}
\author{Francesco Di Renzo}
\affiliation{Universit\`{a} di Parma and INFN gruppo collegato di Parma, 
Viale G.P. Usberti n.7/A - 43124 - Parma, Italy.}
\author{Abhishek Mukherjee}
\affiliation{ECT$^\star$ / FBK, strada delle tabarelle 286 - 38123 - Trento, Italy.}
\affiliation{LISC  / FBK, via sommarive 18 - 38123 - Trento, Italy.}
\author{Luigi Scorzato}
\affiliation{ECT$^\star$ / FBK, strada delle tabarelle 286 - 38123 - Trento, Italy.}
\affiliation{LISC  / FBK, via sommarive 18 - 38123 - Trento, Italy.}

\begin{abstract}
We present the first practical Monte Carlo calculations of the recently proposed Lefschetz thimble formulation of
quantum field theories.  Our results provide strong evidence that the numerical sign problem that afflicts Monte
Carlo calculations of models with complex actions can be softened significantly by changing the domain of
integration to the Lefschetz thimble or approximations thereof.  We study the interacting complex scalar field
theory (relativistic Bose gas) in lattices of size up to $8^4$ using a computationally inexpensive approximation of
the Lefschetz thimble.  Our results are in excellent agreement with known results.  We show that---at least in the
case of the relativistic Bose gas---the thimble can be systematically approached and the remaining residual phase
leads to a much more tractable sign problem (if at all) than the original formulation.  This is especially
encouraging in view of the wide applicability---in principle---of our method to quantum field theories with a sign
problem.  We believe that this opens up new possibilities for accurate Monte Carlo calculations in strongly
interacting systems of sizes much larger that previously possible.
\end{abstract}

\maketitle

{\em Introduction ---} Many important physical systems are characterized by complex actions, when formulated in
terms of a path integral.  But, if the action $S$ is not real, then $e^{-S}$ is not positive semi-definite and it
cannot be interpreted as a probability distribution.  In these cases, Monte Carlo calculations are not applicable
directly.  This is the so called {\em sign problem}.  Many techniques have been proposed to overcome this problem,
with important partial successes, but the sign problem is still unsolved for a variety of important physical
systems and parameter values, such as lattice QCD at high baryonic density \cite{Aarts:2013bla}, or with a
$\theta$-vacuum \cite{VicariPR}, real-time quantum field theories \cite{Berges:2006xc}, the electron structure
calculations \cite{PhysRevLett.71.1148-egas,charutz:4495-egas,baer:6219}, the repulsive Hubbard model
\cite{PhysRevB.41.9301-hubbard}, the nuclear shell model \cite{Koonin19971-shell} or polymer theory
\cite{vilgis2000polymer}, to mention only some of the most famous problems.  In this context, any new idea that
could improve our chances to simulate any of these models on larger lattices than are feasible today would be
extremely valuable.

In a previous work \cite{Cristoforetti:2012su, Cristoforetti:2012uv}, we argued that it may be possible to control
the sign problem by reformulating the associated quantum field theory on a Lefschetz thimble. The Lefschetz
thimble, associated with a saddle point $\phi$, is defined as the hypersurface formed by the union of all paths of
steepest descent (SD) of the complex action ending in that saddle point $\phi$.  Both the Lefschetz thimble and the
saddle point are constructed in an enlarged space obtained by complexifying each field component.  We showed that,
in many cases of interest, this reformulation has the same symmetries and perturbation theory as the original
theory \cite{Cristoforetti:2012su}. Thereafter, appealing to universality we argued that the reformulation has the
same physical content as the original theory.

The benefit of this reformulation is that the action on the Lefschetz thimble has a constant imaginary part, which
can be set to zero without any loss of generality. Thus $e^{-\Re \{S\}}$ can now be interpreted as a probability
distribution in Monte Carlo sampling. Since, the Lefschetz thimble defines a curved integration domain, there can,
in principle, be an additional \emph{residual phase} coming from the Jacobian of the transformation.  However, we
will argue later that this \emph{residual phase}, if at all present, will result in a very mild growth of
stochastic noise.

In this work, we apply our method to the interacting complex scalar field theory describing a relativistic Bose gas
at finite chemical potential. This model is one of the simplest non-trivial examples whose sign problem shares many
features with the more complex systems mentioned above.  Also, in common with lattice QCD, it displays the Silver
Blaze phenomenon \cite{Cohen:2003kd}, i.e., the independence of the physics on the chemical potential up to some
(finite) critical value.  This feature is not accessible to standard Monte Carlo treatments due to the sign
problem.  Quite importantly, this model has been solved through alternative methods \cite{Endres:2006xu,
  Aarts:2008wh, Gattringer:2012df, Gattringer:2012ap}, and as such provides the ideal test bed for new methods,
like ours, for studying the physics of strongly interacting systems.

In the context of Monte Carlo methods, modifications of the domain of integration had been proposed already in
\cite{rom:8241,baer:6219,PhysRevLett.91.150201}.  But those deformations were limited to shifts of the contour in
the imaginary direction.  For many relevant theories, including those considered in \cite{Cristoforetti:2012su},
the shift is zero, and more general transformations are necessary, to reduce the sign problem.  Morse theory
\cite{Pham1983,Witten:2010cx,Cristoforetti:2012su} identifies the Lefschetz thimbles as the appropriate contours of
integration in the more general cases.

{\em Formulation of the model on a Lefschetz thimble ---} The model is defined by the following continuum action:
\begin{equation}
\label{eq:Scont}
S= \int d^4 x
[
|\partial \phi|^2
+
(m^2-\mu^2) |\phi|^2
+
\mu j_0
+
\lambda |\phi|^4
],
\end{equation}
where $\phi(x)$ is a complex scalar field, $j_{\nu} := \phi^*\partial_{\nu} \phi - \phi\partial_{\nu} \phi^*$ and
$\mu$ is the chemical potential.  In this model (as in QCD) the density $\langle n\rangle=\frac{1}{V}\partial \ln Z
/ \partial\mu$ is expected to be zero up to a critical point.  But, this phase transition is hidden in the standard
Monte Carlo method because of the strong sign problem which appears as soon as $\mu \neq 0$.

To formulate and simulate the relativistic Bose gas on a Lefschetz thimble \cite{Cristoforetti:2012su}, we need to
discretize the system defined by Eq.~(\ref{eq:Scont}) and extend the action $S$ holomorphically. This is done by
complexifying both the real and imaginary part of the original complex fields $\phi_x =
\frac{1}{\sqrt{2}}(\phi_{1,x} + i \phi_{2,x})$, as $\phi_{a,x} = \phi^{(R)}_{a,x} + i \phi^{(I)}_{a,x}$, $a=1,2$,
which leads to the action in $d$ dimensions \cite{Aarts:2009hn}:
\begin{eqnarray}
S[\{\phi_{a,x}\}]=
&&
\sum_x
  \left[
    \left(d+\frac{m^2}{2}\right)\sum_a \phi_{a,x}^2
    +\frac{\lambda}{4}(\sum_a \phi_{a,x}^2)^2
\right.\nonumber\\
&&\left.
    -\sum_{a}\sum_{\nu=1}^{d-1}
      \phi_{a,x}\phi_{a,x+\hat{\nu}}
      +\sum_{a,b}
      i\sinh\mu\, \varepsilon_{ab}\phi_{a,x}\phi_{b,x+\hat{0}}
\right.\nonumber\\
&&\left.
      - \cosh\mu\, \delta_{a,b} \phi_{a,x}\phi_{b,x+\hat{0}}
  \right],
\label{eq:S-hol}
\end{eqnarray}
($\varepsilon$ is the 2 dimensional anti-symmetric Levi-Civita symbol).  The observables are defined as:
\begin{eqnarray}
\label{eq:Z}
\langle {\cal O} \rangle_0 &=& \frac{1}{Z_0} \int_{\J_0} \; \prod_{a,x} d\phi_{a,x} \; e^{-S[\phi]} {\cal
  O}[\phi], \nonumber \\
Z_0 &=& \int_{\J_0} \; \prod_{a,x} d\phi_{a,x} \; e^{-S[\phi]},
\end{eqnarray}
where the integration domain $\J_0$ is the Lefschetz thimble \cite{Pham1983,Witten:2010cx} attached to $\phi_{\rm
  glob}$. The configuration $\phi_{\rm glob}$ is the global minimum of the real part of the action $S_R=\Re \{S\}$,
when restricted to the original domain $\mathbb{R}^{2V}$.  More precisely, $\J_0$ is the manifold of real dimension
$N=2V$, defined as union of all the curves of SD for $S_R$, i.e., the curves that are solutions of:
\begin{eqnarray}
\label{eq:SD}
\frac{d}{d\tau} \phi^{(R)}_{a,x}(\tau) &=& - \frac{\delta S_R[\phi(\tau)]}{\delta \phi^{(R)}_{a,x}},
\;\;\;\; \forall a,x,\nonumber \\
\frac{d}{d\tau} \phi^{(I)}_{a,x}(\tau) &=& - \frac{\delta S_R[\phi(\tau)]}{\delta \phi^{(I)}_{a,x}},
\;\;\;\; \forall a,x,
\end{eqnarray}
and that end in $\phi_{\rm glob}$ for $\tau \to \infty$.  

In presence of spontaneous symmetry breaking (SSB), the global minimum $\phi_{\rm glob}$ is degenerate.  But, the
whole procedure can be defined by introducing an explicit term of symmetry breaking: $h \sum_{x,a} \phi_{x,a}$,
where $h$ is a real constant, that selects a specific minimum \cite{Cristoforetti:2012uv} (that can be computed
also analytically).  Since $h$ is real, the global minimum $\phi_{\rm glob}$ of $S_R$ is also a stationary point of
the imaginary part of the action $S_I$, and hence the thimble is well defined.  Physical results are obtained by
extrapolating to $h \to 0$. (In principle, one could define a thimble without introducing $h$ and treat the
symmetries as suggested in \cite{Witten:2010cx}.  This is well suited to local gauge symmetries, but it makes the
study of SSB less clean.)

{\em Aurora Monte Carlo algorithm for sampling the thimble ---} It is possible to generate field configurations on
the Lefschetz thimble with weights given by $e^{-S_R}$ with the help of Langevin dynamics using an algorithm
described in \cite{Cristoforetti:2012su, Cristoforetti:2012uv}, that we review here.  First, let us assume to know
a starting configuration $\phi \in {\cal J}_0$, together with a set of configurations $\phi(k\Delta \tau) \in {\cal
  J}_0$, with $k=1,\ldots,N_{\tau}$, that represent the path of SD connecting $\phi=\phi(0)$ with the configuration
$\phi(\tau=\Delta \tau N_{\tau}$).  Let us assume that $\phi(\tau)$ is sufficiently close to $\phi_{\rm glob}$, so
that the action $S$ can be approximated by its quadratic expansion around $\phi_{\rm glob}$.  Second, we generate a
Gaussian noise $\eta_j(0)$, where $j=1\ldots 2N$ is a multi-index that stands for $(R/I,a,x)$, 
%%%
we evolve it according to:
\begin{equation}
\label{eq:LieSD}
\frac{d}{ds} \eta_{j}(s) = -\sum_{k} \eta_{k}(s) \partial_{k}\partial_{j} S_R[\phi(s)]_{k,j},
\end{equation}
and we project the endpoint with:
\begin{equation}
\eta^{\perp}_j = P_{j,k} \eta_k(\tau), \qquad
\end{equation}
where the $2N\times 2N$ matrix $P$ of rank $N$ is defined in terms of the Hessian matrix $H$ as:
\begin{equation}
P = \frac{H}{\sqrt{H^2}} -1,  \qquad \mbox{ and } \qquad H = \partial^2 S_R[\phi_{\rm glob}].
\end{equation}
Then we normalize the noise vector as:
\begin{equation}
\label{eq:normnoise}
\eta'(\tau) =  r \frac{\eta^{\perp}}{||\eta^{\perp}||},
\end{equation}
where $r$ is a random number distributed according to the $N$-dimensional $\chi$ distribution.  This produces a
Gaussian noise on the tangent space to ${\cal J}_0$ computed in $\phi_{\rm glob}$.  We call such linear vector
space ${\cal G}_0$.  Third, we transport the noise from $s=\tau$ along the path of steepest \emph{ascent} (SA)
to $s=0$ by integrating the ordinary differential equation (ODE):
\begin{equation}
\label{eq:Lie}
\frac{d}{ds} \eta_{j}'(s) = \sum_{k} \eta_{k}'(s) \partial_{k}\partial_{j} S_R[\phi(s)]_{k,j},
\end{equation}
%%%
This ensures that the noise remains tangent to ${\cal J}_0$.  Fourth, we use the
evolved noise $\eta'(0)$ to generate a new configuration as:
\[
\phi_j' = \phi_j - \Delta t\, \frac{\delta S_R[\phi]}{\delta \phi_j}  + \sqrt{2 \Delta t} \, \eta_j'.
\]
In the limit $\Delta t \to 0$ this simulates Langevin dynamics on the thimble.  For $\Delta t > 0$, $\phi'(0)$ will
move away from the thimble of order $(\Delta t)^2$.  To correct this, the fifth step consists in following the path
of SD from $\phi'(0)$ for a length $\tau$ leading to the configuration $\phi'(\tau)$.  Assuming that the action at
$\phi'(\tau)$ can be approximated with its quadratic part (otherwise, we extend $\tau$), we ensure that
$\phi'(\tau)$ belongs to the thimble by projecting it as $\phi(\tau)^{(\rm new)} = P \phi'(\tau)$.  Finally, we
follow the path of SA from $\phi(\tau)^{(\rm new)}$ for a length $\tau$.  The resulting $\phi(0)^{(\rm new)}$ is
the new configuration\footnote{Note that this procedure is not inherently stable, as the one in
  \cite{Cristoforetti:2012su}, but relies on the (verifiable) fact that the integration in $\tau$ always brings
  sufficiently close to the saddle point.}.

The computation of the projector $P$ is done, once for all, at the beginning of the simulation.  However, it must
be applied at every iteration.  This can be done most efficiently in Fourier space, where $H$ and $P$ are diagonal,
although, for this first exploratory study on small lattices, we did not take advantage of this possibility.

The cost of the algorithm depends significantly on the length $\tau$, that should be large enough to stretch out to
the region where the quadratic approximation of the action is good.  But how good is good enough?  We certainly do
not need to constrain the system on the thimble exactly, but only to the extent that the domain of integration
preserves the homology class of the thimble and the reweighting with the phase $e^{i S_I}$ is feasible.  

It is then natural to ask whether $\tau=0$ is already sufficient.  This corresponds to integrating the system on
the vector space ${\cal G}_0$ defined above.  In general, ${\cal G}_0$ does not belong to the same homology class
as ${\cal J}_0$, because the directions of steepest ascent for the quadratic part of the action may not, in
general, be directions of convergence for the full action.

However, in our simulations we observed that such divergences, although they do occur as expected, are very rare
(see below).  This suggests that the integration on ${\cal G}_0$, regularized, say, with a mild cutoff, might
already provide a good approximation.  Of course, such a regulator introduces an unknown bias, and the procedure is
meaningful only if the regulator is eventually removed, by approaching the true thimble further.  Next, we present
our results on ${\cal G}_0$, following which we show how the true thimble can be systematically approached.

{\em Numerical results on ${\cal G}_0$ ---} As discussed above, the simulations on ${\cal G}_0$ are meaningful only
with a regulator.  Instead of introducing an explicit cut of the domain, we regularized by discarding those
simulations that diverged within the observed histories (i.e. $4\times 10^6$ trajectories for $V=4^4$, $10^6$
trajectories for $V=6^4$ and $8 \times 10^5$ trajectories for $V=8^4$).  This procedure introduces an unknown bias,
that can only be removed by approaching the thimble further.  However, the fact that the divergences are very rare
makes the regularization rather unambiguous.  If we consider a common span of the first $8\times 10^5$
trajectories, a divergence occurred with probability $\sim 1.8\%$ on the lattices $V=4^4$, with probability $\sim
0.8\%$ on $V=6^4$, and less than $0.7\%$ on $V=8^4$ ($h=5\times 10^{-3}$).  The results obtained in this way agree
perfectly (within the rather small errors) with the results obtained with the algorithm of \cite{Gattringer:2012df}
and \cite{Aarts:2008wh}\footnote{We thank Gert Aarts, Christof Gattringer and Thomas Kloiber for sharing their
  partially unpublished results with us.}.  In particular, they show the correct scaling with the volume.  Note
that, since ${\cal G}_0$ is a flat manifold, the {\em residual phase}, discussed in \cite{Cristoforetti:2012su,
  Cristoforetti:2012uv} is absent.

\begin{figure}[ht]
\includegraphics[width=\columnwidth]{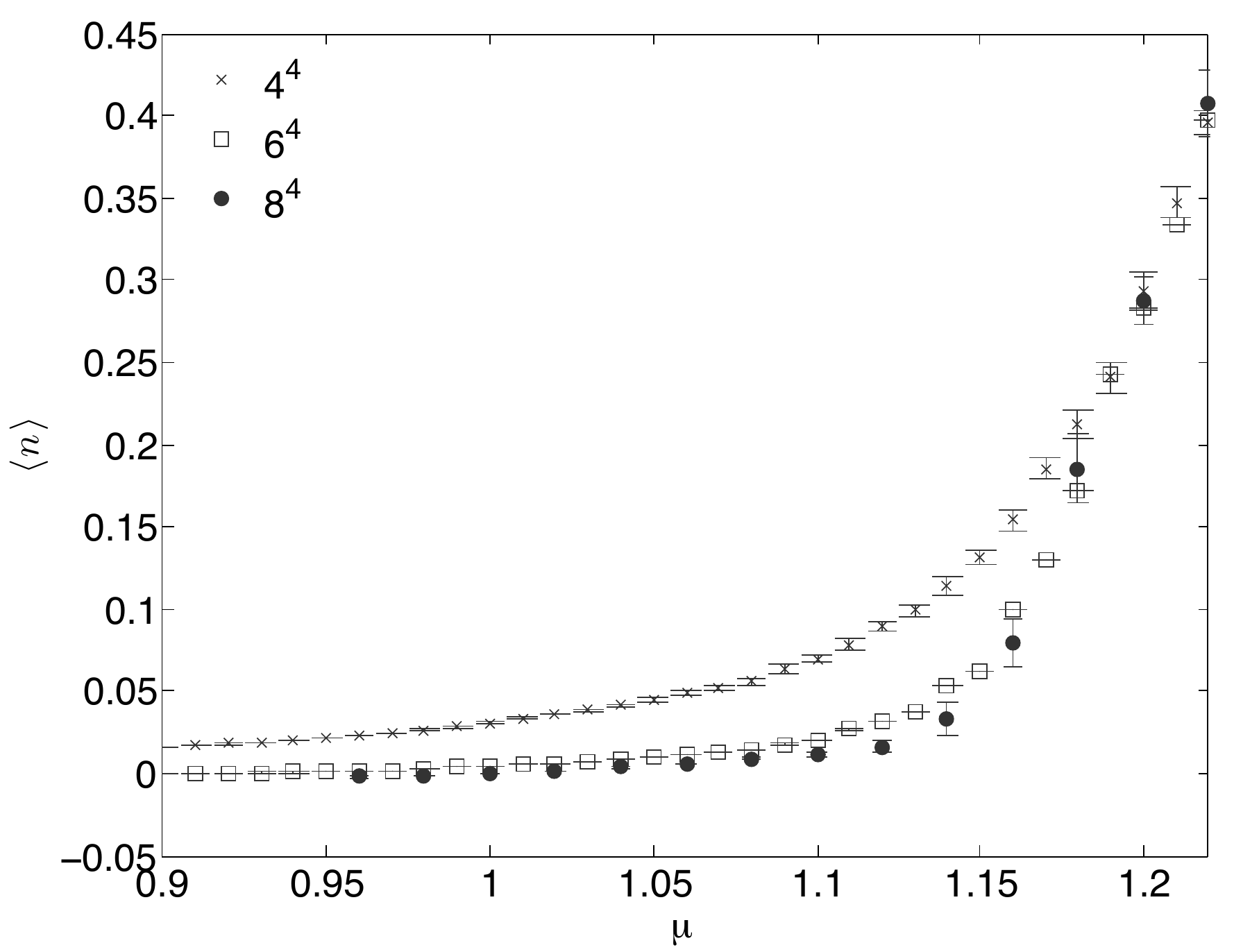}
\caption{\label{fig:1} Average density $\langle n\rangle$ in the critical region for the lattices $V=4^4, 6^4,
  8^4$.}
\end{figure}

\begin{figure}[ht]
\includegraphics[width=\columnwidth]{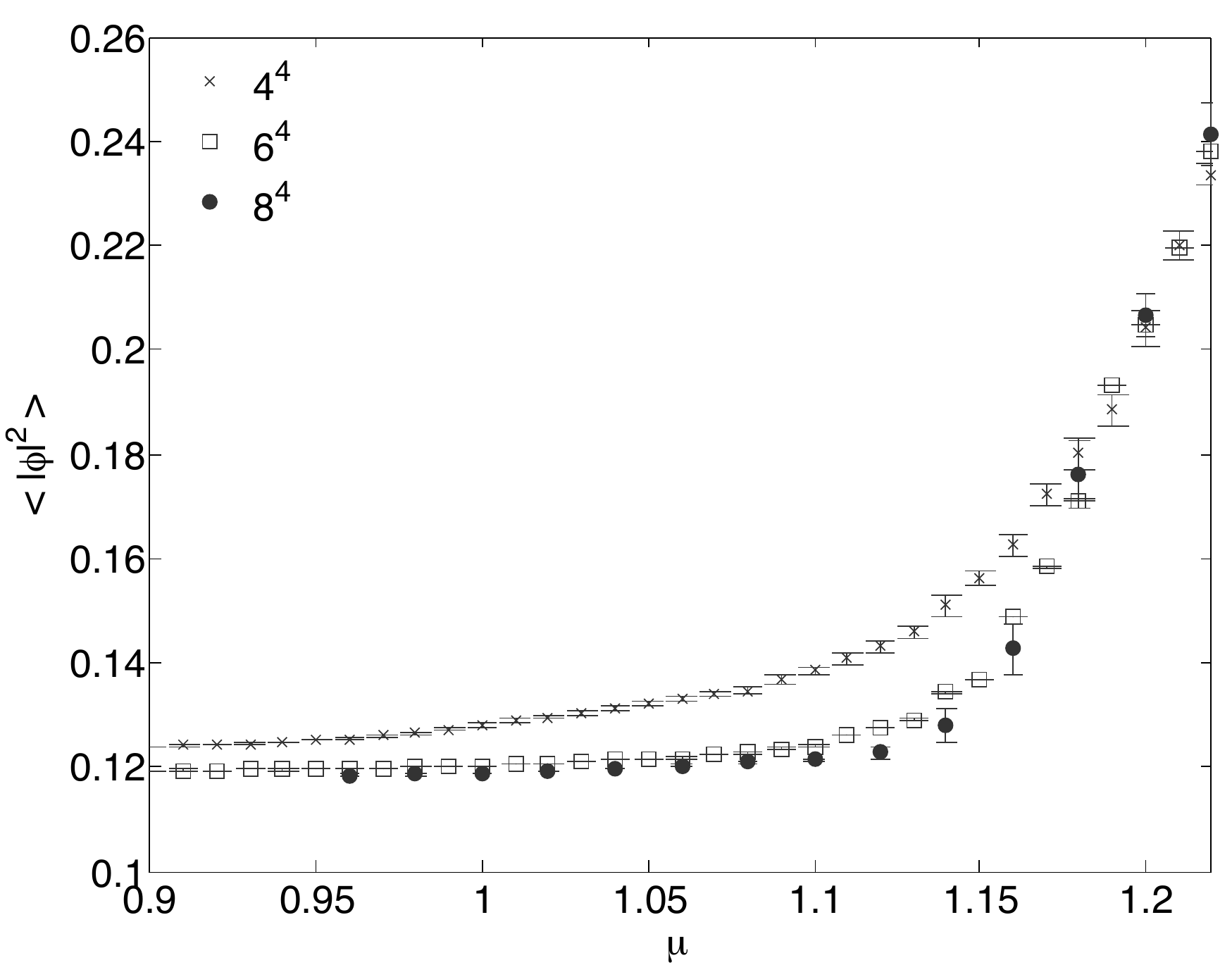}
\caption{\label{fig:2} Same as in Fig.~\ref{fig:1} for the observable $\langle |\phi|^2 \rangle$.}
\end{figure}

We report the results of simulations for the relativistic Bose gas in $3+1$ dimensions ($d = 4$). The mass and
coupling were fixed at $m = 1 = \lambda$, and $\mu$ was varied from $0$ to $1.3$.  In Fig.~\ref{fig:1} and
\ref{fig:2}, we plot our results for the density $\langle n\rangle$ and $\langle |\phi|^2 \rangle$ in the most
interesting range between $\mu=0.9$ and $\mu=1.22$.  In these figures, we see a clear signal of transition around
$\mu \sim 1.1$.  In all the simulations shown here we used $\Delta t=10^{-4}$, but we performed also some tests
with $\Delta t=10^{-3}$ and $\Delta t=10^{-5}$ and we found no significant difference.  The errorbars on each point
are computed from the standard deviation of 10 to 20 independent histories, in order to take the autocorrelation
effects into account.  We used the sources $h=5\times 10^{-3}$ and $h=10^{-3}$ to extract the limit $h\to 0$.


\begin{figure}[ht]
\includegraphics[width=\columnwidth]{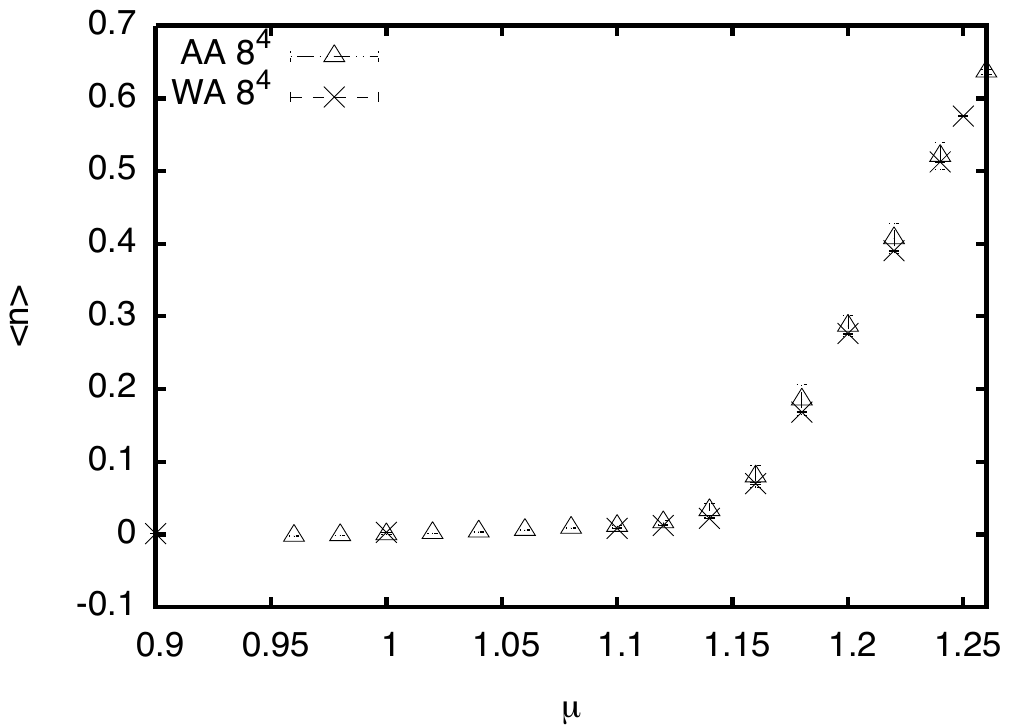}
\caption{\label{fig:3} Comparison of the average density $\langle n\rangle$ obtained with the Worm Algorithm (WA)
  \cite{Gattringer:private} with the Aurora Algorithm (AA) presented here, for the lattice $V=8^4$.  We thank
  C.Gattringer and T.Kloiber for providing us their results.}
\end{figure}

In Fig.~\ref{fig:3} we also compare our results for the average density with those obtained with the algorithm in
\cite{Gattringer:2012df}.

\begin{figure}[ht]
\includegraphics[width=\columnwidth]{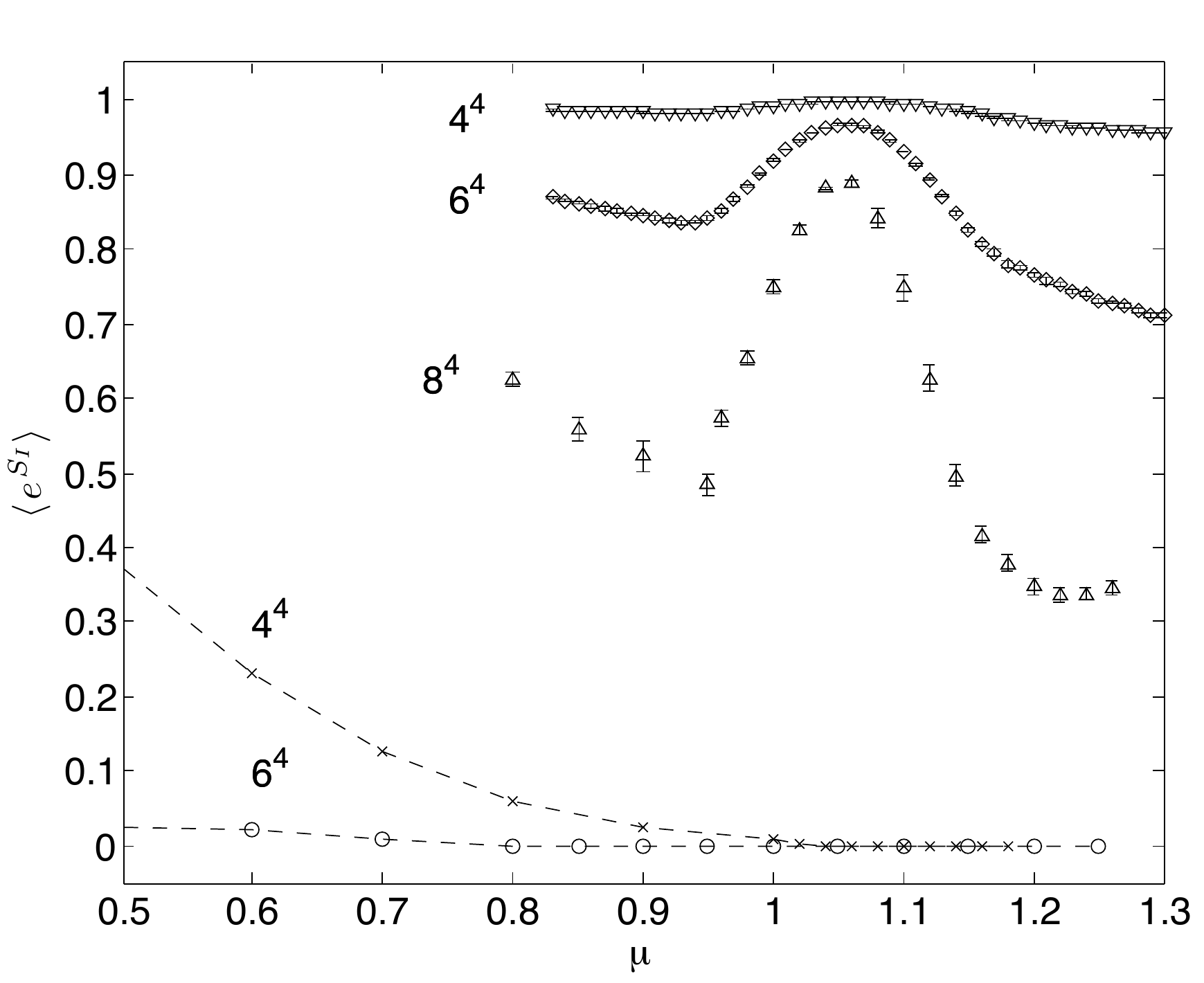}
\caption{\label{fig:fase} The data on the top-right show the average phase obtained with the Aurora algorithm on
  lattices $4^4$, $6^4$ and $8^4$.  It is interesting that the average phase is large precisely in the most
  interesting region just above $\mu=1$.  The dashed lines on the bottom-left display, for comparison, the average
  phase obtained with a naive phase-quenched Monte Carlo algorithm on lattices $4^4$ and $6^4$.  Even on a $4^4$
  lattice, the sign problem in the phase-quenched algorithm, completely hides the interesting region.}
\end{figure}

In Fig.~\ref{fig:fase} we plot the average phase for the same simulations reported above.  The phase is used to
reweight the observables.  However, such reweighting brings corrections to the observables that are unnoticeable,
within the statistical errors.  As expected, the sign problem in ${\cal G}_0$ gradually increases on larger volumes
and moving closer to the thimble will be eventually necessary.
 
{\em Moving closer to the thimble ---} In general, there are two good reasons to move closer to the thimble ${\cal
  J}_0$.  First, to remove the bias introduced by the regulator on $\mathcal{G}_0$.  Second, to keep the sign
problem under control on larger volumes.  However, in the present situation, the divergences are already very rare
and to observe a further measurable reduction would require enormous statistics.  Moreover, the results obtained on
${\cal G}_0$ are already in excellent agreement with the known results, and the reweighting with the phase has no
effect even in the most critical case of the $8^4$ lattice at $\mu=1.2$.  Hence, the results reported here with
$\tau\neq 0$ do not intend to improve the precision of the results obtained above with $\tau= 0$, but rather to
present a first exploration of the feasibility of moving closer to the thimble.

To integrate Eqs.~(\ref{eq:SD}) and (\ref{eq:Lie}), we employed the (classical) 4th order Runge-Kutta method (RK4).
This is an explicit method, that can be used to solve Eq.~(\ref{eq:SD}) and (\ref{eq:Lie}) as initial value
problems (IVP).  We argued in \cite{Cristoforetti:2012su} that, in order to enable a stable integration in the most
general case for large $\tau$, without the need of too tiny $\Delta \tau$, Eq.~(\ref{eq:SD}) should be treated as a
boundary value problem (BVP), by introducing explicit boundary conditions in the neighborhood of the saddle point.
However, it is interesting to see what can be achieved even with the simpler procedure adopted here.  

To evaluate the closeness to the thimble, we monitored the reduction of the fluctuations of the imaginary part of
the action (what really matters for the sign problem), when $\tau$ is increased.  We found that $\tau = 4\times
10^{-2}$ was sufficient to suppress the fluctuations of the imaginary part of the action $S_I$ by a factor $\sim
0.5$ (for $4^4$), a factor $\sim 0.6$ (for $6^4$), and $\sim 0.7$ (for $8^4$).  This test was performed for
$\mu=1.2$, in the critical region. However, a precise integration of the IVP becomes more and more difficult on
increasingly large volumes (the correctness of the integrator can be assessed via reversibility checks).  This
shows that the IVP formulation of the ODE will need to be replaced by a BVP formulation in more difficult
situations.

Finally, note that in this test we neglected the computation of the {\em residual phase} discussed in
\cite{Cristoforetti:2012su, Cristoforetti:2012uv}.  But the excellent agreement with the known results, even
without including the residual phase, supports the idea that its effect is not dramatic and maybe even negligible.

{\em Summary ---} We have reported the first numerical application of the Lefschetz formulation to a nontrivial
model with a hard sign problem.  In particular, we have studied the relativistic Bose gas model at finite chemical
potential.  Our study was restricted to small lattices, but, given the severity of the sign problem, this can be
considered already a very challenging test.  We found excellent agreement with the known results already on the
crudest approximation of the thimble, i.e, the vector space ${\cal G}_0$, once the integral was regulated by
removing the few diverging trajectories.  Moreover, we showed that it is possible to improve the approximation of
the thimble, by following the equations of SD.  

Of course, the sign problem is expected to become worse on larger lattices: moving closer to the thimble will
become more crucial.  Work in progress include developing efficient and stable integration algorithms to achieve a
better approximation of the thimble, study of the scaling for larger system sizes, and application of our method to
other models.

{\em Acknowledgments ---} We are grateful to Christian Torrero for many valuable discussions and to Gert Aarts,
Christof Gattringer and Thomas Kloiber for communications about their results.  This research is partially
supported by the PAT-INFN project AuroraScience, by the EU under Grant Agreement No.~PITN-GA-2009-238353 (ITN
STRONGnet), by INFN through the i.s.~MI11 and by MIUR under contract PRIN2009 (20093BMNPR\_004).  Numerical
simulations were performed with the Aurora prototype at FBK.

\bibliography{../density}{}
\end{document}